\documentclass[preprint,12pt]{elsarticle}

\usepackage{amssymb}
\usepackage{url}

\begin{document}

\begin{frontmatter}



\title{Information transport in multiplex networks}

\author[]{Cunlai Pu\corref{cor1}}
 \ead{pucunlai@njust.edu.cn}
\cortext[cor1]{200 Xiaolingwei, Nanjing 210094, China. Tel: +8613915966537.}
\author[]{Siyuan Li}
\author[]{Xianxia Yang}
\author[]{Jian Yang}
\address{Department of Computer Science and Engineering, Nanjing University of Science and Technology, Nanjing 210094, China}

\begin{abstract}
In this paper, we study  information transport in multiplex networks comprised of two coupled subnetworks. The upper subnetwork, called the logical layer, employs the shortest paths protocol to determine the logical paths for packets transmission, while the lower subnetwork acts as the physical layer, in which packets are delivered by the biased random walk mechanism characterized with a parameter $\alpha$. Through simulation, we obtain the optimal $\alpha$ corresponding to the maximum network lifetime and the maximum number of the arrival packets. Assortative coupling is better than the random coupling and the disassortative coupling, since it achieves much better transmission performances. Generally, the more homogeneous the lower subnetwork, the better the transmission performances are, which is opposite for the upper subnetwork. Finally, we propose an attack centrality for nodes based on the topological information of both subnetworks, and further investigate the transmission performances under targeted attacks. Our work helps to understand the spreading and robustness issues of multiplex networks and provides some clues about the designing of more efficient and robust routing architectures in communication systems.
\end{abstract}


\end{frontmatter}


\section{Introduction}

Communication infrastructures, such as the Internet, mobile phone networks, wireless sensor networks etc., become increasingly interconnected, forming large while still growing interdependent communication networks. For example, in our daily life people can chat with each other through mobile phone networks, or they can communicate through the Internet by using the communication software, such as email, Facebook, etc.  Usually the communication flows will travel across many communication networks, if the distances between sources and destinations are large. In the battlefield, the soldiers and vehicles are connected by different kinds of communication channels (eg. wireless or not) for reliable communication. In all these systems, nodes are connected by more than one kind of links. Each kind of link forms a single network layer, and the network layers are often coupled for specific purposes. Beside the technological systems,  multiple links or interactions among units also widely appear in social and biological systems \cite{Dorogo}. The multiplexity nature of these systems attracts great attention from research community in the past few years, and brings about a new research topic, namely multiplex networks \cite{Boccaletti, Lee}.

Most recently, many researchers focus on the redefinition of the basic structure measures coming from single complex network \cite{Newman09}, and investigate how the multiplexity of layered network structures affect the  dynamical processes such as epidemic spreading \cite{Granell,Buono,Massaro}, percolation \cite{Cellai,Baxter,AZIMI}, games \cite{Gomez,santos,jin}, synchronization \cite{yuan,Tang}, etc. The transmission of information is a common spreading process on communication networks, which is widely studied in single complex networks \cite{Chen,Yang387,Yang83,Wu2008,cunlai13,cunlai12}. The most concerned problem is  how to alleviate the traffic congestion in order to enhance the network capacity. Arenas et al \cite{Arenas} illustrated the phase transition of the traffic from the perspective of critical packet generating rate, which was thereafter used as the indicator of the network capacity. Zhao et al \cite{Zhao05} theoretically discussed the network capacity of several traditional topological structures by considering the delivery capacity as well as the betweenness of the nodes. Most of the other researchers proposed many hard or soft strategies to improve the network capacity \cite{Chen}. The hard strategies are modifying the network structure or optimizing the network resources allocation. For example, Liu et al \cite{Liu07} found that it is effective to improve the network performance by deliberately removing a small set of edges among core nodes. Similarly, Zhang et al \cite{Zhang07} obtained that deleting some edges among high-betweenness nodes will improve the network capacity. Gong et al \cite{Gong08} obtained that for a given network and a routing strategy, there is an optimal resource allocation scheme corresponding to the maximum network capacity which is  inversely proportional to the average length of the routing paths. 

The hard strategies are too expensive or not applicable in real situation. More attention has been paid to the designing of efficient routing strategies \cite{Chen}. In the representative shortest path protocol, traffic flows are directed to pass by the core nodes in order to fastly reach the destinations,  but this will lead to  the traffic congestion. Yan et al \cite{Yan06} proposed an efficient routing strategy, which intentionally avoids traffic flows passing by the core nodes and redistributes the traffic load from core nodes to the marginal nodes. This routing strategy greatly improves the network capacity when the delivery capacity of nodes is identical. Danila et al \cite{Danila} employed a heuristic algorithm to minimize the maximum node betweenness for improving the network capacity. Wang et al \cite{Wangwenxu06} proposed a dynamic routing strategy, which considers both the degrees and loads of the neighbor nodes. There are many other efficient routing strategies which consider the waiting time of packets in the queue \cite{Wangd}, the distance of  the destinations from the neighbor nodes \cite{Chenzy, Echenique}, and so on.

Until very recently, some researchers studied the traffic dynamics on multilayer complex networks. Zhuo et al \cite{Zhuo11} studied the traffic dynamics on two-layer complex networks comprised of a logical layer and a physical layer, and found that the physical layer is more critical to the overall network capacity. Du et al \cite{du14} studied the traffic dynamics on coupled spatial networks with different travel cost among different network layers. Tan et al \cite{Tan14} investigated the effects of interconnections on traffic congestion in two coupled scale-free networks generated by the Barab\'{a}si-Albert (BA) model \cite{barabasi99}. They found that assortative coupling can alleviate traffic congestion much further than disassortative coupling and random coupling when node capacity is allocated according to node usage probability. Nian et al \cite{Nian} compared several routing strategies on two layer degree-coupled networks. Zhou et al \cite{Zhou13} studied the routing issues on multilayered communication networks comprised of a wired network and a wireless network.

In many communication networks, the nodes have a limited power supply, and they can not work when the batteries run out. Especially in wireless sensor networks, the nodes have limited delivery capacity, power, storage capacity, and communication distances, and the most important thing is to make the networks survive as long as possible. Thus, in the power-limited networks, network lifetime \cite{Dietrich,Shah,Chang00} is more important than network capacity. In this paper, we study the information transmission in multiplex networks comprised of two coupled subnetworks. We focus on network lifetime as well as number of arrival packets from the perspective of coupling strategies, routing protocols, network structures, and network attacks.

\section{The network model}
In our multiplex networks, there are two network layers. The two network layers have the same number of nodes and the same average node degree. The  topological structures of the network layers are generated by network models including the  Erd\"{o}s and R\'{e}nyi (ER) model \cite{Erdos} and the static (ST) model \cite{Goh01}. The ER model generates the random networks with a Poisson degree distribution. The static model generates the scale-free networks with a power-law degree distribution. Nodes in different layers are connected based on one of the three rules: random coupling, assortative coupling, and disassortative coupling.  For  the random coupling, nodes in different layers are randomly connected. For the assortative coupling, large-degree nodes in one layer are also connected with the large-degree nodes in the other layer. For the disassortative coupling, large-degree nodes in one layer are connected with small-degree nodes in the other layer. 
From another point of view, each node of the multiplex network has two substitutes, one is in the upper layer, and the other is in the lower layer. The two substitutes are connected with an edge.
\section{The traffic model}
The upper layer in our multiplex network is  the logical layer. In this layer, each time every node generates the packets with  rate $\rho$. If $\rho$ is 1.5, then a node will generate one packet and another one with probability 0.5. The destinations of the packets are randomly selected among the other nodes except the source nodes. The routing paths for the packets are calculated by the shortest path protocol based on the  topological structure of the upper layer. After the routing paths are available, the packets will be transferred to the counterparts of the source nodes in the lower layer for delivery. 

The lower layer  is the physical layer for packets delivery. In the lower layer, packets are stored in the queue of nodes. The queue length is infinite. Each time a node can deliver at most $C$ packets. For simplification purposes, we assume that $C$ is infinite to avoid traffic congestion in  packets transmission. When a node wants to deliver a packet, it will check if the destination of the packet is among its neighbor nodes. If yes, then the packet will be delivered to the destination node directly, and then be transferred to the counterpart  in the upper layer immediately. Otherwise, the packet will be routed by using the biased random walks strategy. Assume there is a packet at node $s$, then the probability that the packet is sent to the $i$th neighbor of node $s$ is as follow:
\begin{eqnarray}
P_{si}=\frac{k_{i}^{\alpha}}{\sum_{j=1}^{k_{s}}k_{j}^{\alpha}},
\end{eqnarray}
where $k_{i}$ is the degree of the $i$th neighbor of node $s$. $\alpha$ is a control parameter. $\alpha=0$ means that the packet will be delivered to all the neighbors with equal probability. When $\alpha > 0$, the packet will be sent to the large-degree nodes with  large probability. When $\alpha< 0$, the small-degree nodes have a large probability to receive the packet. It needs mentioning that all the nodes in the routing path calculated in the upper layer should be orderly visited by the random walk in the lower layer. We assume that in the lower layer each node has $E$ units of energy at the beginning. A node consumes one unit of the energy when delivering a packet. We assume that if any node  uses up its energy,  the multiplex network dies. Then,  the network lifetime $T$ and the  total number of arrival packets $Q$ are two representative properties of the packets transmission.

\begin{figure}
 \centering
\includegraphics[width=5in,height=3in]{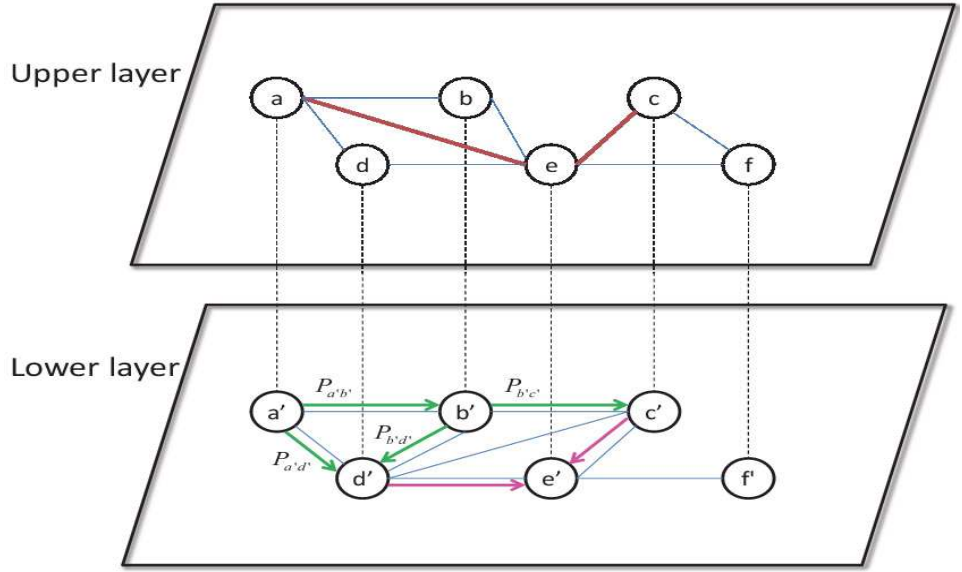}
\caption{Illustration of our traffic model. The multiplex network is comprised of a logical (upper) layer and a physical (lower) layer. A node has a substitute in each of the two layers. The two substitutes of a node are connected with an edge marked with dotted line ($a$-$a'$, $b$-$b'$, $c$-$c'$, $\dots$). In the upper layer, the routing path between any two nodes is calculated with the shortest path algorithm. If there are many shortest paths between two nodes, we randomly choose one of them. For instance, the shortest path between node $a$ and node $c$ is $a-e-c$ marked with red (online). 	In the lower layer, the packets are delivered with the biased random walk mechanism (Eq. 1). For example, there are three possible delivery paths from $a'$ to $e'$ which are $a'-d'-e'$  with probability $P_{a'd'} $, $a'-b'-d'-e'$ with probability $P_{a'b'}P_{b'd'}$, and $a'-b'-c'-e'$ with probability $P_{a'b'}P_{b'c'}$.  Note that in our definition of the random walk, both $P_{d'e'}$ and $P_{c'e'}$ equal 1. When delivering a packet, a node consumes one unit of the energy.
}
\end{figure}

\section{The attack centrality}
For monolayer networks, there are many centrality measures \cite{Newman09} for nodes such as node degree, closeness, betweenness, eigenvector centrality, PageRank centrality, and these measures can be extended to multiplex networks \cite{Boccaletti}. Here, we propose a tunable attack centrality  based on the degrees of nodes in each layer. The  tunable attack centrality of node $i$ is as follows:
\begin{eqnarray}
z_i=k_{up,i}^{\beta}k_{low,i}^{1-\beta},
\end{eqnarray}
where $k_{up, i}$ is the upper-layer degree of node $i$, and $k_{low, i}$ is the lower-layer degree of node $i$. $\beta$ is a free parameter which can adjust the relative importance of $k_{up, i}$ and $k_{ low,i}$ to $z_i$. 
When $\beta=0$, the centrality value of a node is dependent on only its lower-layer degree, while when $\beta=1$, the centrality value of a node equals to its upper-layer degree.
With this tunable attack centrality, we can explore how the packets delivery process affected by the network targeted attacks.

\section{Simulation results}
First, we study the influence of the biased random walks on the packets transmission on multiplex networks. Specifically, we generate each layer of the multiplex networks with the static model. The number of nodes of each layer is $N=1000$ with average node degree $\langle k \rangle=6$. The power-law parameter of each layer is $\gamma=2.5$. The logical layer and the physical layer of the multiplex network are interconnected with the one of the three coupling strategies including randomly coupling, assortative coupling and disassortative coupling. We adjust the control parameter $\alpha$ of the biased random walks in the physical layer, and record the corresponding network properties such as the network lifetime $T$ and the total number of arrival packets $Q$. The results are shown in Fig. 2, which are the average of 1000 independent runs. We see that both $T$ and $Q$ increase with $\alpha$ first, and then decrease with $\alpha$. In the lower layer, when $\alpha >0$, the packets visit large-degree nodes more often, and this leads to a fast energy consumption for large-degree nodes, which have a relatively small population in the scale-free network. When $\alpha<0$, the packets are prone to visit small-degree nodes. Although this alleviates the traffic loads of large-degree nodes, it will make the packets travel much longer in the lower layer, which means more energy will be used in general. The optimal $\alpha$ is negative, where T reaches the peak, as shown in Fig. 2.
For all the coupling strategies, $T$ and $Q$ are positively correlated, and generally the longer the network survives, the more packets the network delivers. Assortative coupling is the best since it achieves better performances especially for $Q$ than the random coupling and disassortative coupling. 
Disassortative coupling is the worst especially for the measure of $Q$ as shown in Fig. 2 (f).
Based on the shortest path protocol, the large-degree nodes  have a large probability to be on the routing paths of packets. On the other hand, according to the random walks theory \cite{Noh04}, the large-degree nodes are more prone to be visited by random walkers compared to the small-degree nodes. Thus, for the assortative coupling, the packets generally arrive at the destinations more early since the large-degree lower layer nodes  they often visit happen to be in the routing paths, which is opposite for  disassortative coupling.
\begin{figure}
\centering
\includegraphics[width=5in,height=2.5in]{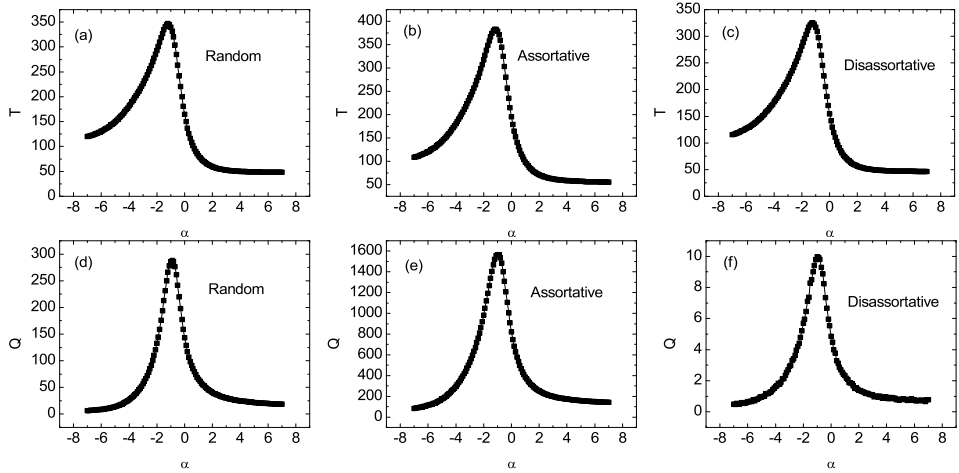}
\caption{Lifetime $T$ and number of delivered packets $Q$ vs. parameter $\alpha$ of the biased random walks for random coupling, assortative coupling and disassortative coupling. Each layer of the multiplex network is generated by the ST model with $N=1000$, $\langle k \rangle=6$, and $\gamma=2.5$. $\rho=0.01$. $E=3000$. Each data point is the average of 1000 independent runs. }
\end{figure}

Then, we study the influence of the network topological structures on the traffic dynamics. We mainly focus on the network heterogeneity which are controlled by the power-law parameter $\gamma$, and adjust the parameter $\gamma$ of each layer. The network size $N$ and the average node degree $\langle k \rangle$ of each layer are fixed to be 1000 and 6 respectively. The random coupling, assortative coupling and disassortative coupling are considered for the interconnection of the upper layer and the lower layer. We show the results of $T$ as a function of $\gamma$, which are the average of 1000 independent runs. In Fig. 3, we see that for random coupling $T$ decreases and converges with increase of $\gamma$ of the upper layer,   when the lower layer is fixed to be either the ER network or the ST network. On the other hand, if we fix the upper layer to be either the ER network or the ST network, $T$ will increase with $\gamma$ of the lower layer. The results of Fig. 3 indicate that for random coupling, to achieve large network lifetime, the topological structure of the upper layer should be as heterogeneous as possible, while the topological structure of the lower layer should be as homogeneous as possible. The reason is that a heterogeneous upper layer  results in a small average routing path, which means generally the packets will stay  for a short time in the multiplex network and require a relatively small amount of energy for their delivery. 
However, for the lower layer the homogeneous network structure leads to a relatively balanced load distribution,  which increases the network lifetime.
The results for the assortative coupling (Fig. 4) and disassortative coupling (Fig. 5) are similar to that of random coupling. There is only an exception for assortative coupling (Fig. 4 (a)), when the lower layer is fixed to be the ER network. $T$ slightly increases with $\gamma$, and reaches the peak value around $\gamma=3$, and then decreases with $\gamma$. 
The reason for this exception is that, for assortative coupling the large-degree upper layer nodes also have large-degree counterparts in the in the lower layer. When the heterogeneity of the upper layer begins to decrease, in the lower ER network loads of large-degree  nodes are reduced accordingly,  which results in a direct increase of the network lifetime. When the network structure of the upper layer becomes more homogeneous, the average routing path is increased accordingly, which leads to a consumption of more energy and a decrease of the network lifetime.
\begin{figure}
\centering
\includegraphics[width=5in,height=2.5in]{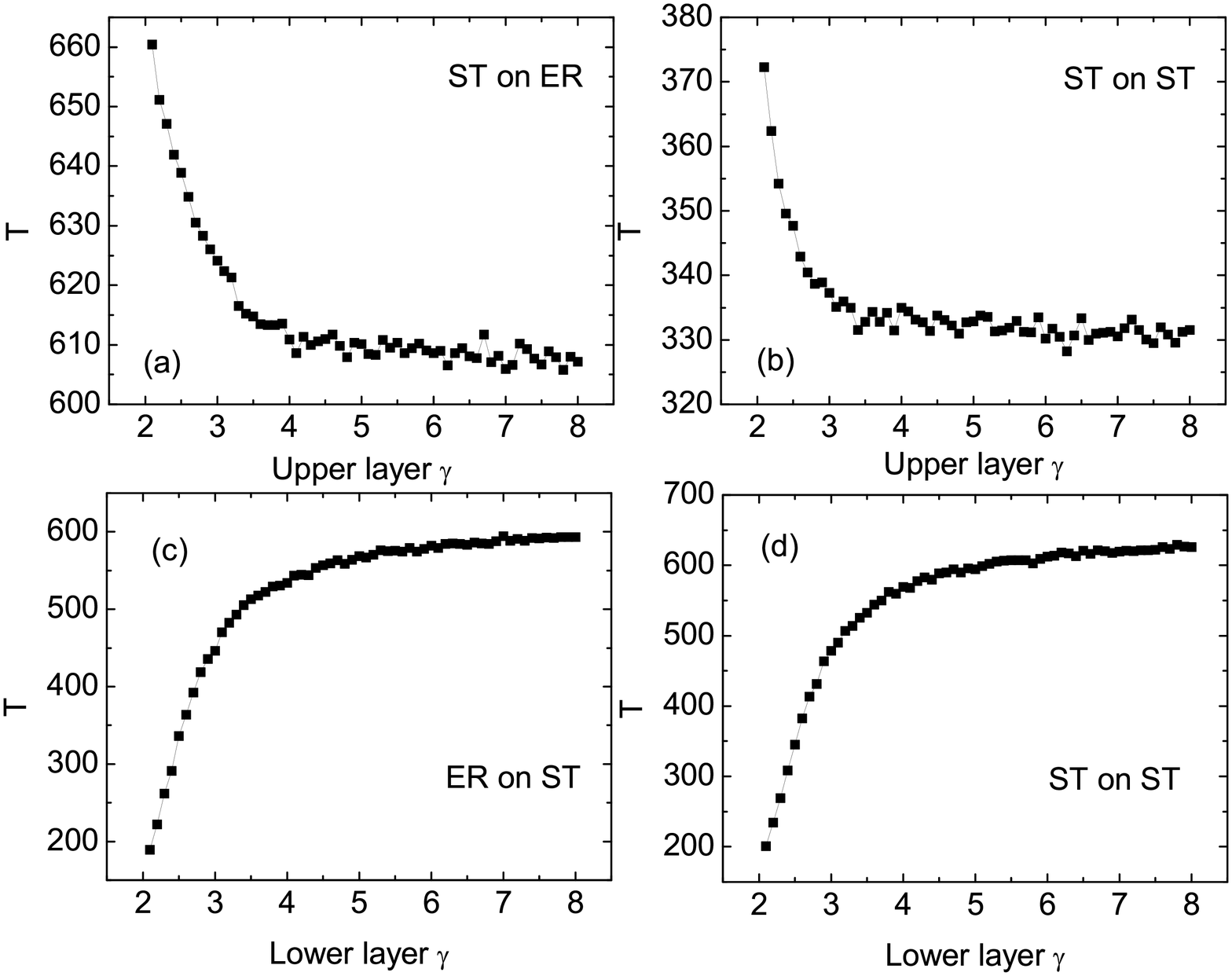}
\caption{Network lifetime $T$ vs. power-law parameter $\gamma$ for the random coupling strategy. ST on ER means that the upper layer is the ST network and the lower layer is the ER network. ST on ST means that both the upper layer and the lower layer are the ST networks. ER on ST represents that the upper layer is the ER network, and the lower layer is the ST network. For (a) and (b) the lower layer is fixed, while for (c) and (d) the upper layer is fixed. Each layer of the multiplex networks have $N=1000$ nodes with average node degree $\langle k \rangle=6$. $\rho=0.01$. $E=3000$. The power-law parameter of the fixed ST networks in (b) and (d) is $\gamma=2.5$. The results are the average of 1000 independent runs.}
\end{figure}
\begin{figure}
\centering
\includegraphics[width=5in,height=2.5in]{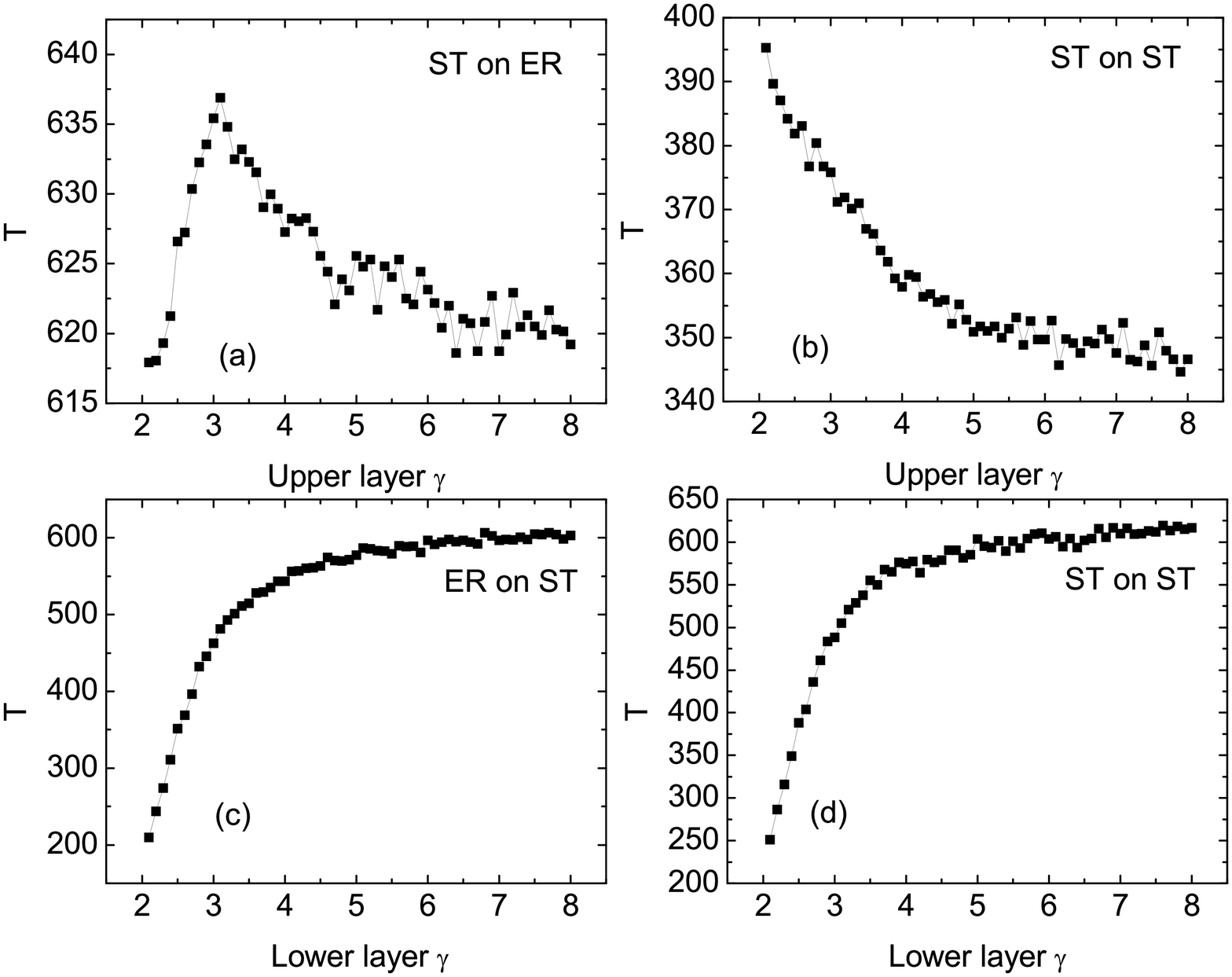}
\caption{Network lifetime $T$ vs. power-law parameter $\gamma$ for the assortative coupling strategy. The multiplex networks and the parameters are the same as in the Fig. 3. The results are the average of 1000 independent runs.}
\end{figure}
\begin{figure}
\centering
\includegraphics[width=5in,height=2.5in]{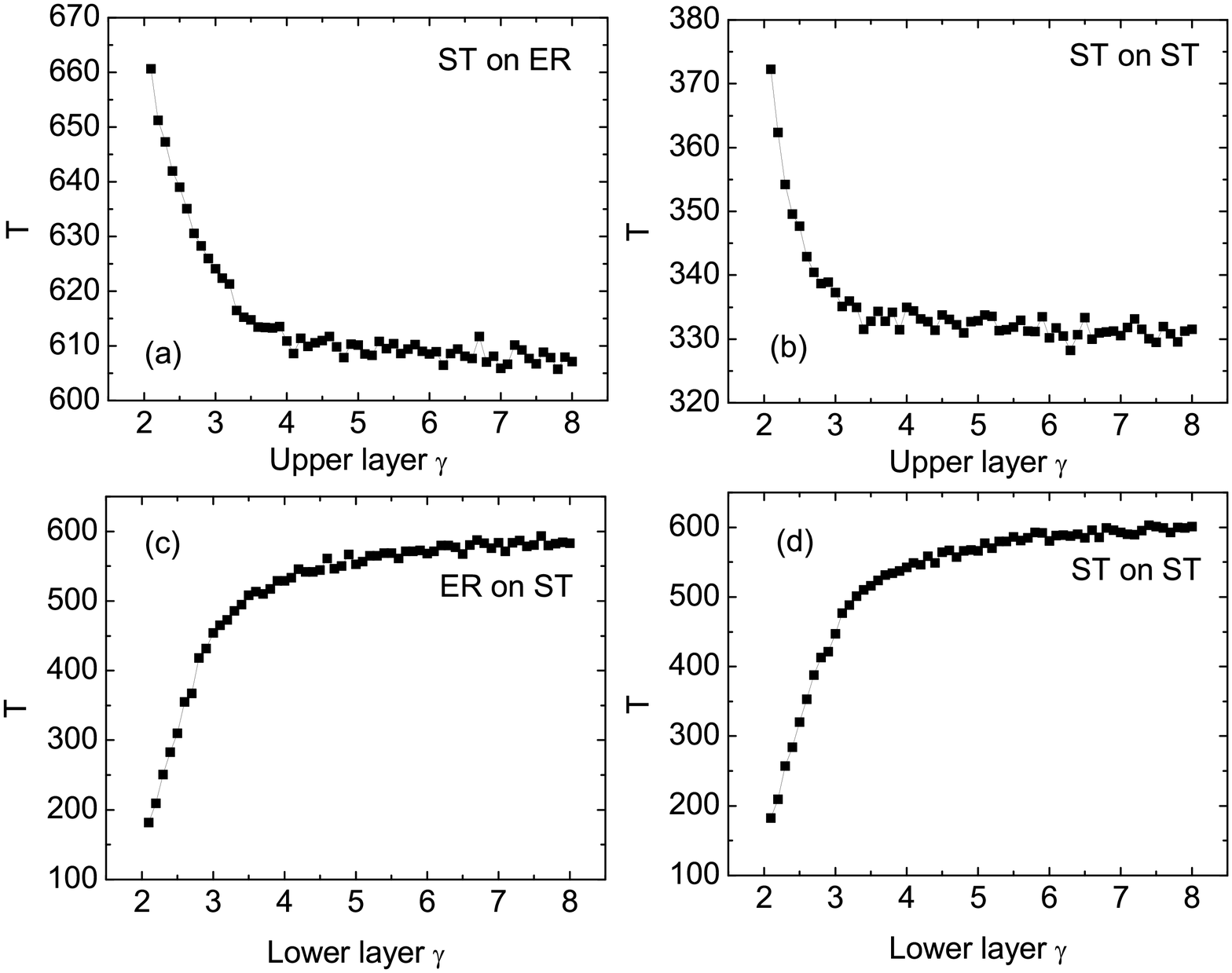}
\caption{Network lifetime $T$ vs. power-law parameter $\gamma$ for the disassortative coupling strategy. The multiplex networks and the parameters are the same as in the Fig. 3. The results are the average of 1000 independent runs.}
\end{figure}

Finally, we study the traffic dynamics in multiplex networks under network disturbance such as the targeted attacks. The attack centrality of each node is calculated with Eq. 2. In the attack process, we rank all the nodes based on their centrality values and attack the nodes with the largest centrality. It is worth remarking that when we attack a node, we remove its counterparts and the incident edges from both the physical layer and the logical layer. In the simulation, we mainly focus on how the transmission performances such as $T$ and $Q$ affected by the parameter $\beta$ of the attack centrality measure. Both the upper layer and the lower layer are generated by the ST model. The network parameters of each layer is $N=1000$, $\langle k \rangle=6$ and $\gamma=2.5$. The fraction of nodes we attack is set to be 30\%. The parameter $\alpha$ of the biased random walks is set to be 0. The initial amount of energy $E$ for each node in the physical layer is 6000. The simulation results are shown in Fig. 6, where each data point is the average of 1000 independent runs. We see that the results for different coupling strategies are very different. For random coupling, $T$ increases first, and then decreases with $\beta$, while $Q$ decreases first, and then increases with $\beta$. 
Those nodes with large upper-layer degrees or large lower-layer degrees are the critical nodes for $T$ and $Q$. These critical nodes deliver much more packets than the other nodes, and usually use up their energy much early than the others. Thus, removing these critical nodes will generally result in a more balanced use of energy, which will increase the network lifetime. On the other hand, removing these critical nodes on average increases the delivery path length and the difficulty of arriving at the destinations for the packets, and this leads to the decrease of arrival packets. There is a good compromise between attacking the nodes with large upper-layer degrees and the nodes with large lower-layer degrees when $\beta$ is around 0.5, where $T$ and $Q$ reach the extreme values, as shown in Fig. 6. 
For assortative coupling, both $T$ and $Q$ are not sensitive to $\beta$. For disassortative coupling, there is a phase transition around $\beta=0.5$. For all the three coupling methods, there are jumps of $T$ and $Q$ when $\beta=1$.

 To explain these results, we measure the variation of node rankings based on the attack centrality for different $\beta$. We denote the ranking of a node $i$ (which has a counterpart in each of the two layer) by $r_i$, which is a function of $\beta$. We define that the average change of node rankings is as follows:
\begin{eqnarray}
\langle \Delta r \rangle=\frac{\sum_{i=1}^{N}(r_i(\beta)-r_i(0))}{N},
\end{eqnarray}
where $N$ is the number of nodes in the multiplex network which equals the number of nodes in each of the two layers. We show the results of $\langle \Delta r \rangle$ in Fig. 7, where the multiplex networks and the parameters are the same as in Fig. 6. Each data point is the average of 1000 independent runs. For random coupling, $\langle \Delta r \rangle$ increases with $\beta$ continuously, which means the node rankings changes continuously with $\beta$. For assortative coupling, $\langle \Delta r \rangle$ is fixed to zero, which indicates that the node rankings do not change, and the attacked nodes are actually the same for different $\beta$. For disassortative coupling, there is a phase transition of $\langle \Delta r \rangle$ around $\beta=0.5$, which indicates that the node rankings are constant but different for $\beta<0.5$ and $\beta>0.5$.
For all the three coupling methods, there are jumps of $\langle \Delta r \rangle$ at $\beta=1$, where the attack centrality values are calculated by only the topological information of the upper layer.
\begin{figure}
\centering
\includegraphics[width=5in,height=2.5in]{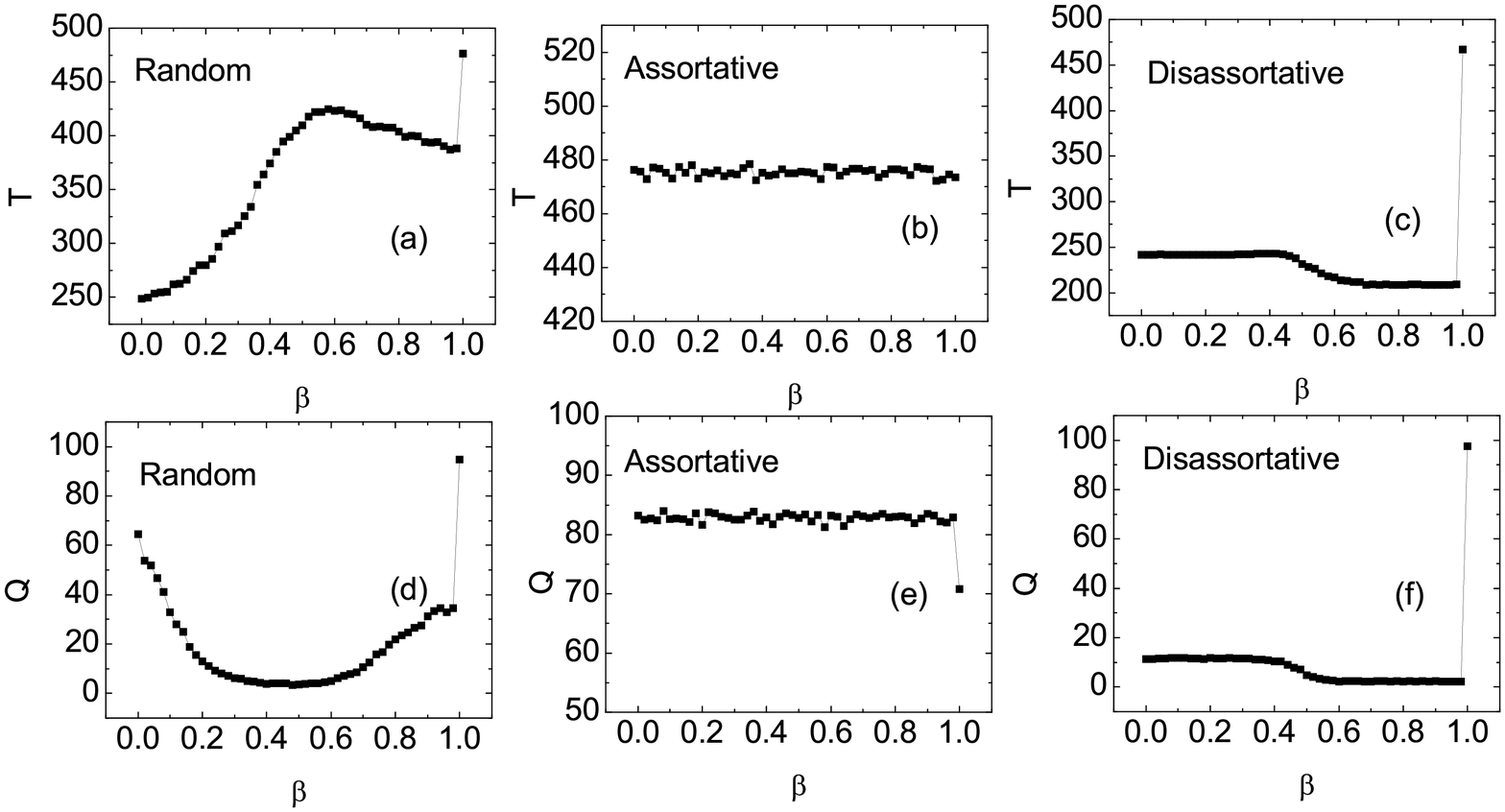}
\caption{Network lifetime $T$ and number of arrival packets $Q$ vs parameter $\beta$ of the attack centrality measure. In the targeted attacks, 30\% nodes with the largest centrality are attacked. Each layer of the multiplex network is generated by the ST model with $N=1000$, $\langle k \rangle=6$, and $\gamma=2.5$. $\rho=0.01$. $E=6000$. Each data point is the average of 1000 independent runs.}
\end{figure}
\begin{figure}
\centering
\includegraphics[width=3in,height=2.5in]{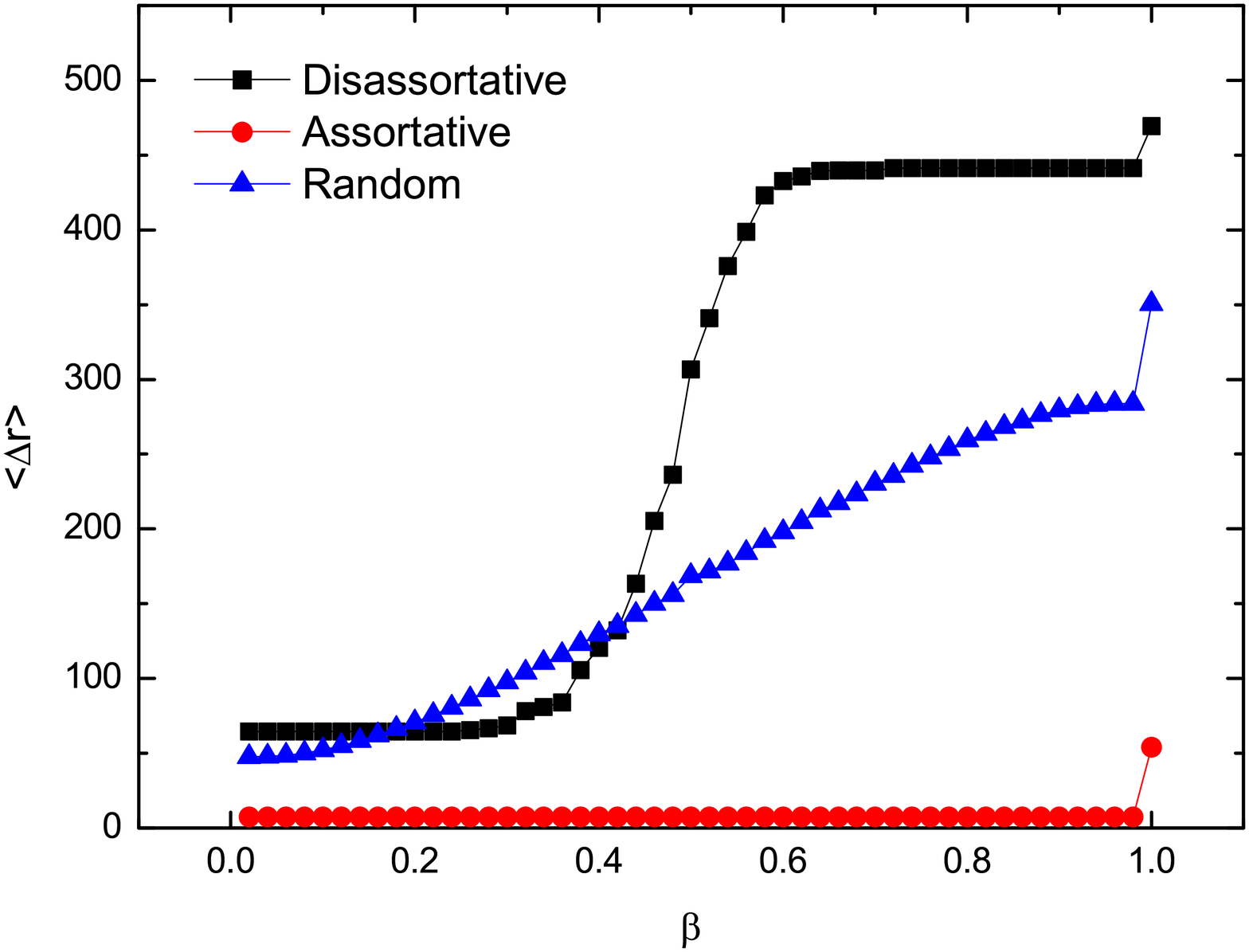}
\caption{$\langle \Delta r \rangle$ vs. $\beta$ for different coupling strategies. The multiplex networks and the related parameters are the same as in Fig. 6. The results are the average of 1000 independent runs.}
\end{figure}

\section{Conclusions}
Multiplex networks theory is an elegant tool for investigating the communication networks which are essentially interconnected with each other. In this paper, we study the traffic dynamics embedded in the multiplex networks. In previous research of traffic dynamics, the network capacity of complex networks is widely discussed. However, we focus on the network lifetime and the number of arrival packets, which are more critical than network capacity in power-limited communication networks such as the wireless sensor networks. 

In our model, the multiplex networks consists a logical network layer and a physical network layer. In the logical layer, the routing paths of packets are computed by the shortest path algorithm. In the physical layer, each node is power-limited, and packets are transmitted by the biased random walk mechanism controlled by a parameter $\alpha$. 

We obtain the optimal parameters $\alpha_{opt}$ leading to the largest network lifetime and the largest number of arrival packets respectively. 
We find that assortative coupling of the logical layer and the physical layer is much better than the random coupling and the disassortative coupling since it achieves much better and more robust network performances than the other two coupling strategies. Generally, the more heterogeneous the logical layer, the better the transmission performances are, which is opposite for the physical layer. 

Finally, we propose a tunable attack centrality with a control parameter $\beta$ for nodes in multiplex networks, based on which we further study how the transmission performances are influenced by the targeted attacks on nodes. Simulation results demonstrate that the changes of the network lifetime and the number of arrival packets with parameter $\beta$ are very different for different coupling strategies. Generally, our attack centrality is more viable in multiplex networks with random coupling and disassortative coupling.

\section*{Acknowledgments}
This work was  supported by the National Natural Science Foundation of China (Grant No. 61304154), the Specialized Research Fund for the Doctoral Program of Higher Education of China  (Grant No. 20133219120032), and the Postdoctoral Science Foundation of China (Grant No. 2013M541673).




\end{document}